\documentclass[aps,nofootinbib,preprintnumbers,amsmath,amssymb]{revtex4}
\usepackage{latexsym}
\usepackage{epsfig}
\def\dirac#1{\rlap{\slash}#1}
\begin{document}
%
%\topmargin -1cm
%\oddsidemargin -0.8cm
%\evensidemargin -0.8cm

%\preprint{IFT-P.0xx/2005}
\title{Double Higgs
Production and Quadratic Divergence Cancellation in Little Higgs
Models with T-Parity}
\author{Claudio O. Dib}
  \email{claudio.dib@usm.cl}
\affiliation{Department of Physics, Universidad T\'ecnica Federico
Santa Mar\'{\i}a, Valpara\'{\i}so, Chile}
\author{Rogerio Rosenfeld}
   \email{rosenfel@ift.unesp.br}
\affiliation{Instituto de F\'{\i}sica Te\'orica - UNESP,
 Rua Pamplona, 145, 01405-900 S\~ao Paulo, SP, Brasil}
\author{Alfonso Zerwekh}
    \email{alfonsozerwekh@uach.cl}
\affiliation{Department of Physics, Universidad Austral de Chile,
 Vald\'{\i}via, Chile}

\date{July 2005}

\begin{abstract}
We analyze double Higgs boson production at the Large Hadron Collider in the context
of Little Higgs models. In double Higgs production, the diagrams
involved are directly related to those that cause the cancellation
of the quadratic divergence of the Higgs self-energy, providing
a robust prediction for this class of models. We find
that in extensions of this model with the inclusion of a so-called
T-parity, there is a significant enhancement in the cross sections
as compared to the Standard Model.
\end{abstract}

\maketitle

\section{Introduction}

The presence of quadratic divergences in loop corrections to the
scalar Higgs boson self-energy is responsible for the so-called
hierarchy problem of the Standard Model (SM); namely, there is no
natural way of having a ``light'' mass (i.e. $\sim 10^2$ GeV) for
the Higgs given that loop corrections induce contributions to the
mass of the order of the GUT scale --or in general the high energy
scale above which new physics enters and the SM ceases to be an
effective theory. In supersymmetric extensions of the SM this
problem is absent since the bosons are protected from quadratic
divergences by their relation to supersymmetric (fermion)
partners\cite{susy1}. The hierarchy problem is also absent in
models in which the electroweak symmetry is dynamically broken,
since the scalar particles are not fundamental but composite in
those cases \cite{techni2}.

Recently a new kind of model was proposed which can solve the
hierarchy problem of the scalar Higgs boson. In these models,
called Little Higgs (LH) models \cite{little3}, the Higgs boson is
a pseudo-Goldstone boson and its mass is protected by a global
symmetry and, unlike supersymmetry, quadratic divergence
cancellations are due to contributions from new particles with the
same spin.

The phenomenology of these models has been discussed with respect
to indirect effects on precision measurements \cite{precision4}
and direct production of the new particles introduced
\cite{pheno5}.

Since these early contributions, several variations in the LH
framework have been proposed \cite{review6}. However, the
cancellation of quadratic divergences is inherent to any LH model
and this requires definite relations among certain couplings.
Therefore, any process that involves exclusively these couplings
is a robust prediction of the LH mechanism regardless of model
variations.

In this article we study a process that has ingrained in it the
cancellation of quadratic divergences of top-quark loops, namely
double Higgs production. In section II we review the model and the
derivation of the masses and couplings of the relevant particles
by appropriate diagonalization of mass matrices and show the
cancellation of quadratic divergences directly in the broken
symmetry phase. In section III we derive the amplitudes for double
Higgs production at the Large Hadron Collider (LHC) and compute the cross sections. Our
results are presented in section IV and we conclude in section V.

\section{Masses, couplings and quadratic divergences in the Littlest Higgs
Model}

There are many variations of Little Higgs models today, which
differ in the symmetry groups and representations of the scalar
multiplets, but they all have in common a mechanism of
cancellation of the quadratic divergence for the mass of the
lightest remaining scalars at one loop order.
After the spontaneous breakdown of a global underlying symmetry at
a scale $4\pi f$ (supposedly not much higher than a few TeV to
avoid fine tuning), the model contains a large multiplet of
pseudo-Goldstone bosons, which includes the SM Higgs doublet.
While most members of the multiplet receive large masses (again,
of a few TeV), the mass of the Higgs boson is protected from
quadratic divergences at one loop, and therefore remains naturally
smaller. The cancellation is related to the existence of an extra
(heavier) top-like quark and its interactions with the scalar
sector, feature which is common to all LH models. Consequently, a
good test to distinguish a little Higgs from other cases should be
based on a signal sensitive to this particular feature of
divergence cancellation and rather insensitive to other features.
The Higgs pair production at LHC is one of such signals, since it
is based on exactly the same diagrams that enter the quadratic
divergence cancellation (Fig.~1), except for the insertion of two
gluons (Fig.~2 and 3).

In order to work out the details, we make use of the Littlest
Higgs model, which is a simple case but contains all the necessary
ingredients. After spontaneous breakdown of the (high-energy)
underlying symmetry, the Little Higgs lagrangian below the scale
$4\pi f$ \cite{littlest7} can be written as a non-linear sigma
model based on a coset $SU(5)/SO(5)$ symmetry:
\begin{equation}
{ \cal L}_{\Sigma} = \frac{1}{2} \frac{f^2}{4}
        {\rm Tr} | {\cal D}_{\mu} \Sigma |^2,
\label{Sigma}
\end{equation}
where the subgroup $[SU(2)\times U(1)]^2$ of $SU(5)$ is promoted
to a local gauge symmetry. The covariant derivative is defined as
\begin{equation}
{\cal D}_\mu \Sigma=  \partial_\mu\Sigma - i \sum_{j=1}^2\left(
g_j( W_j\Sigma +  \Sigma W_j^T) + g'_j (B_j\Sigma + \Sigma B_j^T)
\right).
\end{equation}
To exhibit the interactions, one can expand $\Sigma$  in powers of
${1}/{f}$ around its vacuum expectation value $\Sigma_0$
\begin{equation}
\Sigma = \Sigma_0 + \frac{2 i}{f} \left(
\begin{array}{ccccc}
\phi^{\dagger} & \frac{h^{\dagger}}{\sqrt{2}} &
{\mathbf{0}}_{2\times 2} \\
\frac{h^{*}}{\sqrt{2}} & 0 &
\frac{h}{\sqrt{2}} \\
{\mathbf{0}}_{2\times 2} & \frac{h^{T}}{\sqrt{2}} &
\phi
\end{array} \right)
+ {\cal O}\left(\frac{1}{f^2}\right),
\end{equation}
where $h$ is the doublet that will remain light and $\phi$ is a
triplet under the unbroken $SU(2)$. The non-zero vacuum
expectation value of the field $\langle \Sigma \rangle =\Sigma_0 $
leads to the breaking of the global $SU(5)$ symmetry to $SO(5)$
and also breaks the local gauge symmetry $[SU(2)\times U(1)]^2$
into its diagonal subgroup, which is identified with the standard
model $SU_L(2)\times U_Y(1)$ symmetry group. Following the
notation of Han {\it et al.} \cite{pheno5}, we will denote the
usual standard model gauge bosons mass eigenstates as $W^{\pm}_L$,
$Z_L$ and $A_L$, where the subscript $L$ denotes light in order to
distinguish from the heavy states with mass of order $f$, denoted
by $W^{\pm}_H$, $Z_H$ and $A_H$.

The standard model fermions acquire their masses via the usual
Yukawa interactions. However, in order to cancel  the top quark
quadratic contribution to the Higgs self-energy, a new-vector like
color triplet fermion pair, $\tilde t$ and $\tilde t^{\prime c}$,
with quantum numbers $({\mathbf{3,1}})_{Y_i}$ and
$({\mathbf{\bar{3},1}})_{-Y_i}$ must be introduced. Since they are
vector-like, they are allowed to have a bare mass term which is
{\it chosen} such as to cancel the quadratic divergence above
scale $f$.

The coupling of the standard model top quark to the
pseudo-Goldstone bosons and the heavy colored fermions in the
littlest Higgs model is chosen to be
\begin{equation}
{\cal{L}}_Y = \frac{1}{2}\lambda_1 f \epsilon_{ijk} \epsilon_{xy}
\chi_i \Sigma_{jx} \Sigma_{ky} u^{\prime c}_3 + \lambda_2 f
\tilde{t} \tilde{t}^{\prime c} + {\rm h.c.}, \label{yuk}
\end{equation}
where $\chi_i=(b_3, t_3, \tilde{t})$ and $\epsilon_{ijk}$ and
$\epsilon_{xy}$ are antisymmetric tensors. The new model
parameters $\lambda_1,\ \lambda_2$ are supposed to be of the order

of unity.

The linearized part of Eq.~\ref{yuk} describing the third
generation mass terms and interactions with the neutral Higgs
field, denoted by $h^0$, (before spontaneous symmetry breaking,
SSB), is given by:
\begin{equation}
{\cal L}_{t} = \lambda_2 f \tilde{t} \tilde{t}^{\prime c} - i
\lambda_1 \sqrt{2} t_3 h^0 u_3^{\prime c} +  \lambda_1 f \tilde{t}
u_3^{\prime c} - \frac{\lambda_1}{f} \tilde{t} h^0  h^{0 \ast}
u_3^{\prime c} + h.c.
\end{equation}
After SSB, we write $h^0 = 1/\sqrt{2} (v + H)$, and follow
Perelstein {\it et al.} \cite{PPP8} in defining left handed fields
$t_{3 L} \equiv t_3$, $\tilde{t}_L \equiv \tilde{t}$ and right
handed fields $\bar{u}_{3 R}^{\prime}\equiv u_3^{\prime c}$,
$\bar{\tilde{t}}^{\prime}_{R} \equiv \tilde{t}^{\prime c}$ to
obtain
\begin{eqnarray}
{\cal L}_{t} &=& \left(
\begin{array}{cc} \bar{u}_{3 R}^{\prime} &
\bar{\tilde{t}}^{\prime}_{R}
\end{array}
\right)
\left(
\begin{array}{cc} -i \lambda_1 v & \lambda_1 f (1-v^2/2 f^2) \\
0 & \lambda_2 f
\end{array}
\right)
\left(
\begin{array}{c}  t_{3 L} \\ \tilde{t}_L
\end{array}
\right) - i \lambda_1 H \bar{u}_{3 R}^{\prime}  t_{3 L}  \\
\nonumber
 && - \lambda_1 \frac{v}{f} H \bar{u}_{3 R}^{\prime} \tilde{t}_L -
 \frac{\lambda_1}{2 f} H^2 \bar{u}_{3 R}^{\prime} \tilde{t}_L +
h.c.
\end{eqnarray}

\begin{figure}[htb]
\epsfig{file=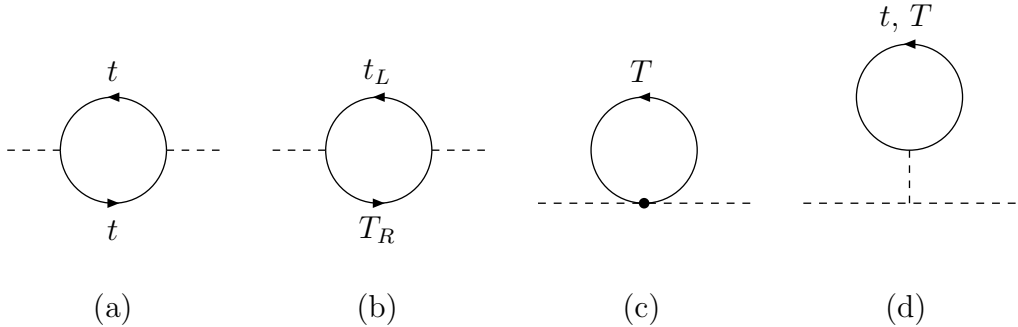}
\caption{One-loop corrections to the Higgs
mass, to order $v/f$: (a) standard top quark loop, (b) mixture of
standard and extra top quark loop, (c) extra top quark loop with a
4-particle vertex, and (d) tadpoles with standard and with extra
top quark loops. There are other diagrams but they are suppressed
by factors of order $(v/f)^2$ or higher.} \label{quadratic}
\end{figure}

In order to leave the fermion mass term in its standard form we
make the field redefinitions $t_{3 L} \rightarrow - i t_{3 L}$ and
$\tilde{t}_L \rightarrow - \tilde{t}_L$  in the left-handed fields
resulting in the following lagrangian:
\begin{eqnarray}
{\cal L}_{t} &=&
- \left( \begin{array}{cc} \bar{u}_{3 R}^{\prime} & \bar{\tilde{t}}^{\prime}_{R}
\end{array} \right)
\left( \begin{array}{cc}  \lambda_1 v & \lambda_1 f (1-v^2/2f^2)
\\ 0 & \lambda_2 f
\end{array} \right)
\left( \begin{array}{c}  t_{3 L} \\ \tilde{t}_L
\end{array} \right)
- \lambda_1 H \bar{u}_{3 R}^{\prime}  t_{3 L}\label{Lag-mass}
\\ \nonumber
 && + \lambda_1 \frac{v}{f} H \bar{u}_{3 R}^{\prime} \tilde{t}_L +
 \frac{\lambda_1}{2 f} H^2 \bar{u}_{3 R}^{\prime} \tilde{t}_L +
h.c.
\end{eqnarray}
Diagonalizing the mass matrix
\begin{equation}
M = \left( \begin{array}{cc}  \lambda_1 v & \lambda_1 f
(1-v^2/2f^2) \\ 0 & \lambda_2 f
\end{array} \right)
\end{equation}
we obtain the usual result for the eigenvalues corresponding to
the top quark $t$ and the heavy top $T$ which are, up to order
${\cal O} (v/f)$:
\begin{equation}
m_t  = \frac{\lambda_1 \lambda_2}{\sqrt{\lambda_1^2 +
\lambda_2^2}}\; v\;; \;\;\;
m_T = \sqrt{\lambda_1^2 +
\lambda_2^2}\; f.
\end{equation}
In our numerical code, we will use as input the values of $m_T$
and $f$, from which one obtains the required values of $\lambda_1$
and $\lambda_2$:
\begin{equation}
\lambda_{1 (2)}^2 = \frac{m_T^2}{2 f^2} \left( 1 + (-) \sqrt{1 - 4
\frac{m_t^2 f^2}{v^2 m_T^2} } \right).
\label{lambdas}
\end{equation}
From equation (\ref{lambdas}), one clearly sees that there is a
condition relating top masses and v.e.v.'s that these models
impose:
\begin{equation}
m_T > 2 \frac{m_t{}v} f \simeq \sqrt{2} f
\end{equation}
and which we incorporate in our analysis. The relevant couplings
between Higgs and top quarks are obtained in a straightforward
manner after diagonalization of Eq.~\ref{Lag-mass} and are given
by (the corresponding vertices are obtained via multiplication by
$(-i)$):
\begin{eqnarray}
 g_{Htt} &=&  \frac{\lambda_1 \lambda_2}{\sqrt{\lambda_1^2 + \lambda_2^2}}
   \\ \nonumber
 g_{HT_Rt_L} &=& \frac{\lambda_1^2}{\sqrt{\lambda_1^2 +
   \lambda_2^2}} \\ \nonumber
 g_{HTT} &=&-\frac{\lambda_1^2\lambda_2^2}{(\lambda_1^2 +
   \lambda_2^2)^{3/2}}  \; \frac{v}{f}  \\ \nonumber
 g_{HHTT} &=& -\frac{1}{f} \frac{\lambda_1^2}{ \sqrt{\lambda_1^2
   + \lambda_2^2}} \\ \nonumber
 g_{HHtt} &=&   \frac{m_t}{f^2} \frac{ \lambda_1^2}{\lambda_1^2 +
   \lambda_2^2 - 4 f v'/v^2} \\ \nonumber
 g_{Ht_RT_L} &=& -\frac{\lambda_1 \lambda_2^3}{(\lambda_1^2 +
    \lambda_2^2)^{3/2}} \; \frac{v}{f}
\end{eqnarray}
The relevant Feynman diagrams for the Higgs self-energy are shown
in figure (\ref{quadratic}).

%\begin{center}
%\end{center}

The cancellation of tadpole diagrams requires that
\begin{equation}
g_{Htt} m_t + g_{HTT} m_T = 0
\end{equation}
whereas the cancellation of higgs self-energy quadratic divergences implies
\begin{equation}
g_{Htt}^2 + g_{HTT}^2 + g_{HT_Rt_L}^2 + g_{Ht_RT_L}^2 + g_{HHtt} \;m_t
+ g_{HHTT} \;m_T = 0
\end{equation}

These conditions are satisfied up to terms of order ${\cal O}
(v/f)$ by the masses and  couplings listed
above.\cite{discrepancy89}

In the simplest LH models, strict bounds on the parameters are
obtained. In particular, electroweak precision constraints require
$f > 3.5$ TeV \cite{precision4}. However, in a recent variation on
the littlest Higgs model, where a so-called T-parity that
interchanges the two subgroups $[SU(2)\times U(1)]_1$ and
$[SU(2)\times U(1)]_2$ of $SU(5)$ is introduced, can significantly
lower this bound to $f > 500$ GeV \cite{TParity9}. This is an
important point for the phenomenology of these models, since a
lower $f$ implies larger deviations from the SM.  Since the T-odd
states do not participate in the cancellation of quadratic
divergences, our calculation is valid in models with T-parity as
well. T-parity also forbids the generation of a vacuum expectation
value for the triplet scalar field (i.e., $v'=0$ in the notation
of T.~Han {\it et al.}\cite{pheno5}), which is one of the causes
for easing the electroweak constraints.

\section{Amplitudes for double Higgs production}

We now turn to Higgs boson production at the LHC in the LH model,
which involves the very same couplings responsible for the
cancellation of quadratic divergences.

\begin{figure}[htb]
\epsfig{file=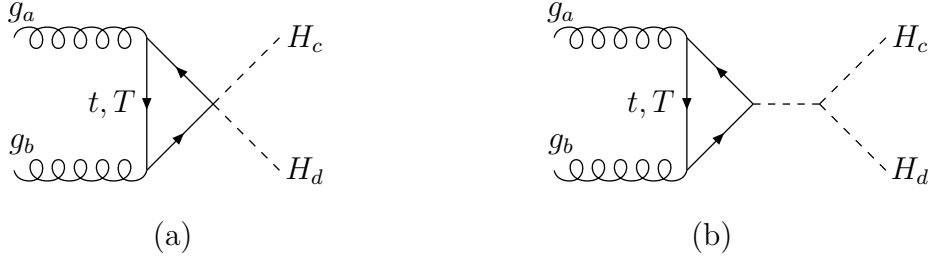} \caption{Triangle contributions to Higgs
boson pair production at LHC in a Little Higgs model.}
\label{triangles}
\end{figure}

\begin{figure}[htb]
\epsfig{file=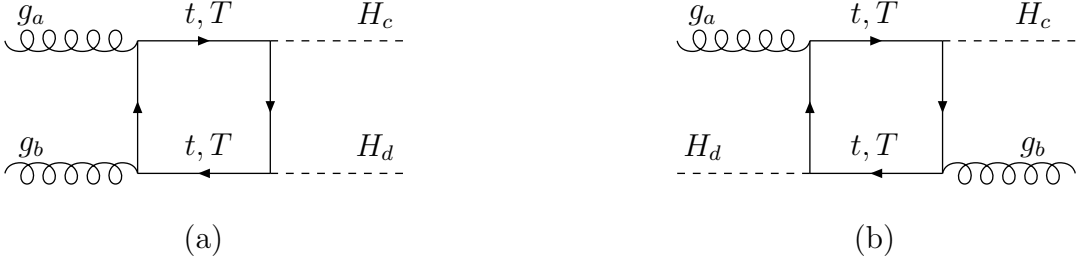} \caption{Box contributions to Higgs boson
pair production at LHC in a Little Higgs model.} \label{boxes}
\end{figure}

Gluon-gluon fusion is the dominant mechanism for SM Higgs boson
pair production at the LHC \cite{dhiggs10}. The amplitude for $gg
\rightarrow HH$ is dominated by top quark loops, in the form of
triangle and box diagrams. Figs.~\ref{triangles} and \ref{boxes}
show the case for a LH model. The SM case is similar,
except that Fig.~\ref{triangles}.a and all extra heavy-top loops are
absent. We also would like to point out that T-parity forbids a term like $h h \phi$ in the radiatively generated Coleman-Weinberg potential. Therefore, there is no contribution of the heavy scalar in Fig.~\ref{triangles}.b and the trilinear Higgs coupling is the same as in the SM.

%\begin{center}
%\end{center}

%\begin{center}
%\end{center}

Let us now write the expressions for the amplitude. In what
follows, the external momenta $p_a$, $p_b$, $p_c$, $p_d$ are
defined as incoming. The contribution from triangle diagrams is
given by:
\begin{eqnarray}
i {\cal M}^\triangle(g_a g_b \rightarrow H_c H_d) &=&
-\frac{\alpha_s}{4 \pi^3} \delta_{AB} \left( g_{Htt} I(m_t)
\frac{g_{HHH}}{\hat{s}-M_H^2- i M_H \Gamma_H} \right.
\\ \nonumber
 & &  \left. +\, g_{HTT} I(m_T) \frac{g_{HHH}}{\hat{s}-M_H^2 -
 i M_H \Gamma_H} + g_{HHtt} I(m_t) +
g_{HHTT} I(m_T)  \right)
\end{eqnarray}
where the integral $I(m_Q)$ is:
\begin{equation}
I(m_Q) = \int d^4 q \; \frac{Tr\left[ (\dirac q +m_Q) \gamma^\mu
(\dirac q +  \dirac p_a + m_Q) \gamma^\nu (\dirac q +  \dirac p_a
+ \dirac p_b + m_Q)\right]}{[q^2 - m_Q^2] [(q+p_a)^2 -
m_Q^2][(q+p_a+p_b)^2 - m_Q^2]}  \epsilon_\mu(p_a)
\epsilon_\nu(p_b).\label{int-triang}
\end{equation}

This integral reduces to the following result:
\begin{equation}
I(m_Q) = i \; 4 \pi^2 \; m_Q [1+ (2 m_Q^2 - \hat{s}/2 )
C_{0}(0,0,\hat{s},m_Q^2,m_Q^2,m_Q^2) ] \epsilon(p_a)  \cdot
\epsilon(p_b) ,\label{PV-triang}
\end{equation}
where $C_{0}$ is the scalar Passarino-Veltman
integral\cite{Passarino} defined as:
\begin{equation}
C_{0} = \int \frac{d^4 q}{i \pi^2}  \;\frac{1}{[q^2 - m_Q^2]
[(q+p_a)^2 - m_Q^2][(q+p_a+p_b)^2 - m_Q^2]}
\end{equation}
The contribution from box diagrams can be written as:
\begin{eqnarray}
i {\cal M}^\Box(g_a g_b \rightarrow H_c H_d) &=&
-\frac{\alpha_s}{8 \pi^3} \delta_{AB} \left( g_{Htt}^2 I_1(m_t) +
g_{HTT}^2 I_1(m_T) +  I_2(m_t,m_T)  \right)
\end{eqnarray}
Because the Higgs vertex is not diagonal in the $(t,T)$ flavor,
there are two types of boxes. The function $I_1$ comes from boxes
with either only standard top quarks or only extra heavy-top quarks in
them, whereas the function $I_2$ comes from boxes with both tops
and extra heavy-tops.

There are two basic box diagrams, planar (Fig.~\ref{boxes}.a) and
non-planar (Fig.\ref{boxes}.b), according to whether the Higgses
are adjacent in the loop or not. A planar box with a single type
of quark is given by:
\begin{eqnarray}
I^P_1(m_Q) &=& \int d^4 q \; \frac{Tr\left[ (\dirac q +m_Q)
\gamma^\mu (\dirac q +  \dirac p_a + m_Q) \gamma^\nu (\dirac q +
\dirac p_a + \dirac p_b + m_Q) (\dirac q +  \dirac p_a + \dirac
p_b + \dirac p_c +m_Q) \right]}{[q^2 - m_Q^2] [(q+p_a)^2 -
m_Q^2][(q+p_a+p_b)^2 - m_Q^2][(q+p_a+p_b+p_c)^2 - m_Q^2]}
 \nonumber \\
 && \times \epsilon_\mu(p_a)\epsilon_\nu(p_b)
%  \nonumber \\ &&
 \quad +\, \Big\{p_a \leftrightarrow p_b\Big\} \, + \,    \Big\{p_c
\leftrightarrow p_d\Big\} \, + \,         \Big\{p_a
\leftrightarrow p_b, \, p_c \leftrightarrow p_d\Big\} ,
\end{eqnarray}
while a non-planar box is:
\begin{eqnarray}
I^{NP}_1(m_Q) &=& \int d^4 q \; \frac{Tr\left[ (\dirac q +m_Q)
\gamma^\mu (\dirac q +  \dirac p_a + m_Q) (\dirac q +  \dirac p_a
+ \dirac p_c + m_Q) \gamma^\nu (\dirac q +  \dirac p_a + \dirac
p_b + \dirac p_c +m_Q) \right]}{[q^2 - m_Q^2] [(q+p_a)^2 -
m_Q^2][(q+p_a+p_c)^2 - m_Q^2][(q+p_a+p_b+p_c)^2 - m_Q^2]}
 \nonumber \\
 && \times \epsilon_\mu(p_a)\epsilon_\nu(p_b) \quad + \, \Big\{
p_a \leftrightarrow p_b\Big\}
\end{eqnarray}
The total contribution for boxes with a single type of quark is
then
\begin{equation}
I_1(m) = I^P(m) + I^{NP}(m)
\end{equation}

We also have to compute the contribution of box diagrams with both
$t$ and $T$ running in the loop. There are also planar and a
non-planar contributions in this case. For the planar contribution
we have:
\begin{eqnarray}
I_2^P(m_t,m_T) &=&\int d^4 q \; \frac{Tr\left[ (\dirac q +m_t) \gamma^\mu
(\dirac q +  \dirac p_a + m_t) \gamma^\nu
(\dirac q +  \dirac p_a + \dirac p_b + m_t) \left(g_{HT_Rt_L}
\frac{1+\gamma_5}{2} + g_{Ht_RT_L}\frac{1-\gamma_5}{2} \right) \right]}
{[q^2 - m_t^2]
[(q+p_a)^2 - m_t^2][(q+p_a+p_b)^2 - m_t^2]}
\nonumber \\
&&
 \times \frac{ \left[(\dirac q +  \dirac p_a + \dirac p_b + \dirac p_c +m_T)
 \left(g_{HT_Rt_L}
 \frac{1-\gamma_5}{2} + g_{Ht_RT_L}\frac{1+\gamma_5}{2} \right)
 \right]}{[(q+p_a+p_b+p_c)^2 - m_T^2]}
 \nonumber \\
 &&\times  \epsilon^\mu(p_a)\epsilon^\nu(p_b)
% +\nonumber \\ &&
\quad +\, \Big\{p_a \leftrightarrow p_b\Big\} \, + \,    \Big\{p_c
\leftrightarrow p_d\Big\} \, + \,  \Big\{p_a \leftrightarrow p_b,
\, p_c \leftrightarrow p_d\Big\}
\end{eqnarray}
and the non-planar contribution is written as:
\begin{eqnarray}
I_2^{NP}(m_t,m_T) &=&\int d^4 q \;  \frac{Tr\left[ (\dirac q +m_t)
\gamma^\mu (\dirac q +  \dirac p_a + m_t) \left(g_{HT_Rt_L}
\frac{1+\gamma_5}{2} + g_{Ht_RT_L}\frac{1-\gamma_5}{2} \right)
(\dirac q +  \dirac p_a + \dirac p_b + m_t) \gamma^\nu  \right]}
{[q^2 - m_t^2] [(q+p_a)^2 - m_t^2][(q+p_a+p_b)^2 - m_t^2]}
 \nonumber \\
 && \times \frac{ \left[(\dirac q +  \dirac p_a + \dirac p_b
+ \dirac p_c +m_T) \left(g_{HT_Rt_L} \frac{1-\gamma_5}{2} +
g_{Ht_RT_L}\frac{1+\gamma_5}{2} \right)
 \right]}{[(q+p_a+p_b+p_c)^2 - m_T^2]}
 \nonumber \\
 && \times \epsilon^\mu(p_a)\epsilon^\nu(p_b) \quad + \,
\Big\{p_a \leftrightarrow p_b\Big\}
\end{eqnarray}
Accordingly, the total contribution for boxes with both $t$ and
$T$ is given by:
\begin{equation}
I_2 = I_2^P(m_t,m_T) + I_2^P(m_T,m_t) + I_2^{NP}(m_t,m_T)  +
I_2^{NP}(m_T,m_t).
\end{equation}

We can then express these integrals in terms of Passarino-Veltman
functions, in a way analogous to Eqs.~(\ref{int-triang}) and
(\ref{PV-triang}).  We computed these transformations using the
package FeynCalc \cite{FeynCalc11}. Since this procedure is
straightforward and the final result is rather long, we did not
include the expressions here.

\section{Cross section results}

Using the total scattering amplitude ${\cal M}(g_a g_b \rightarrow
H_c H_d) = {\cal M}^\triangle + {\cal M}^\Box$ we can build the
partonic differential cross section:
\begin{equation}
\frac{d \hat{\sigma}}{d \Omega} = \frac{1}{128 \pi^2 \hat{s}}
\sqrt{1 - 4 M_H^2/\hat{s}} \;\overline{| {\cal M}|^2}.
\label{diff-cross}
\end{equation}
We must point out that we have included a factor of $1/2$ due to
the identical particles in the final state. Consequently, to
obtain the total cross section one must integrate
Eq.~(\ref{diff-cross}) over the whole $4\pi$ solid angle. Here
$\overline{| {\cal M}|^2}$ is the squared matrix element averaged
over initial color and helicity states:
\begin{equation}
\overline{| {\cal M}|^2} = \frac{1}{32} \sum_{i,j=1,2} | {\cal
M}_{i j }|^2
\end{equation}
where the sum is over the two physical gluon polarizations.

We performed the calculation in the center-of-momentum frame of
the gluons. In that case, the transversality condition of the
gluon polarization vectors also implies $p_a \cdot\epsilon(p_b)=
p_b \cdot \epsilon(p_a) = 0$. Therefore we use the following
parametrization (recalling that all 4-momenta were defined as
incoming):
\begin{eqnarray}
p_a &=& (\sqrt{\hat{s}}/2,0,0,\sqrt{\hat{s}}/2) \\ \nonumber
p_b &=& (\sqrt{\hat{s}}/2,0,0,-\sqrt{\hat{s}}/2) \\ \nonumber
p_c &=& (-\sqrt{\hat{s}}/2,0,-p \sin \theta, -p \cos \theta) \\ \nonumber
p_d &=& (-\sqrt{\hat{s}}/2,0,p \sin \theta, p \cos \theta) \\ \nonumber
\epsilon^{(1)} &=& (0,1,0,0) \\ \nonumber
\epsilon^{(2)} &=& (0,0,1,0)
\end{eqnarray}
where the Higgs boson center-of-mass momentum is $p= \sqrt{
\hat{s}/4 - M_H^2}$. We numerically integrate the
Passarino-Veltman functions using the package LoopTools
\cite{looptools12}. Finally, we obtain the $pp \rightarrow HH$
cross section at the LHC by convoluting the partonic cross section
with the gluon distribution function:
\begin{equation}
\sigma (pp \to HH) = \frac{K}{2}   \int dx_1 dx_2 \; [g_1(x_1,Q^2)
g_2(x_2,Q^2) + g_2(x_1,Q^2) g_1(x_2,Q^2)] \hat{\sigma} (gg \to HH)
\theta (x_1 x_2 s - 4 M_H^2),
\end{equation}
where we used the Set 3 of CTEQ6 leading gluon distribution
function with momentum scale $Q^2 = \hat{s}$ \cite{CTEQ13}. A
$K=2$ factor was included to take into account QCD corrections
\cite{Dawson}.

%For the LHC, with $\sqrt{s} = 14$ TeV we obtain $\sigma (pp \to HH) = 38$ fb for
%$M_H = 120$ GeV. With an expected luminosity of $10^{34}$ cm$^{-2}$ s$^{-1}$
%\cite{lumLHC}  one would have of the order of $4000$ events in one year.

\begin{figure}
\epsfig{file=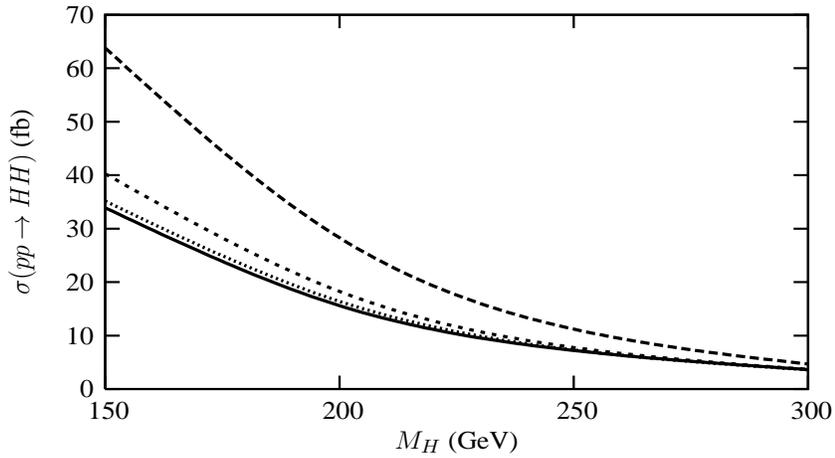,height=6cm, width=11cm} %
\caption{ Cross section for double Higgs production at the LHC for
$M_T = 4$ TeV and $f =500$ GeV (dashed line), $1000$ GeV (short
dashed line) and $2000$ GeV (dotted line). In solid line is shown
the SM result.}\label{result1}
\end{figure}

In Fig.~\ref{result1} we plot the cross section for the double
Higgs production process at the LHC for fixed $M_T = 4$ TeV, a
Higgs boson mass in the range $150$--$300$ GeV and for $f =500$,
$1000$ and $2000$ GeV. As expected, we find that the largest
deviations from the SM result occurs for small Higgs boson mass
and small decay constant $f$. In this sense it is important to
consider models with T-parity, where $f$ is not required to be too
large. Our results are otherwise consistent with the authors of
Ref.~\cite{Jing}, where values around $f=3.5$ TeV are used.

%
%
%
%%-----------FIGURE 4 --------------------------------------

\begin{figure}
\epsfig{file=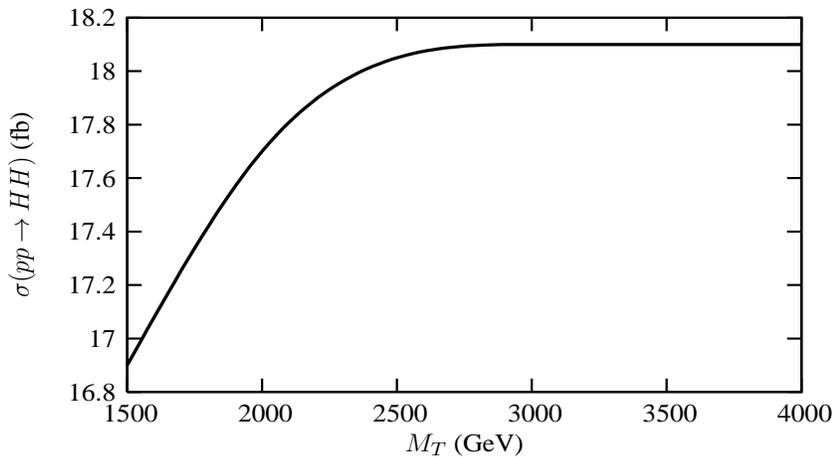,height=6cm, width=11cm} %
\caption{ Cross section for double Higgs production at the LHC for
$M_H = 200$ GeV and $f = 1000$ GeV, as a function of the heavy top
quark mass $M_T$.}\label{result2}
\end{figure}

In order to explore the dependence on the heavy top quark mass we
plot in Fig.~\ref{result2} we plot the cross section for the
double Higgs production process at the LHC for fixed $M_H = 200$
GeV and $f = 1000$ as a function of $M_T$. We can see that the
result grows with $m_T$, reaching a constant limit for $m_T$ above
$\sim 2.5$ TeV.

%
%
%
%%-----------FIGURE 5 --------------------------------------

\begin{figure}
\epsfig{file=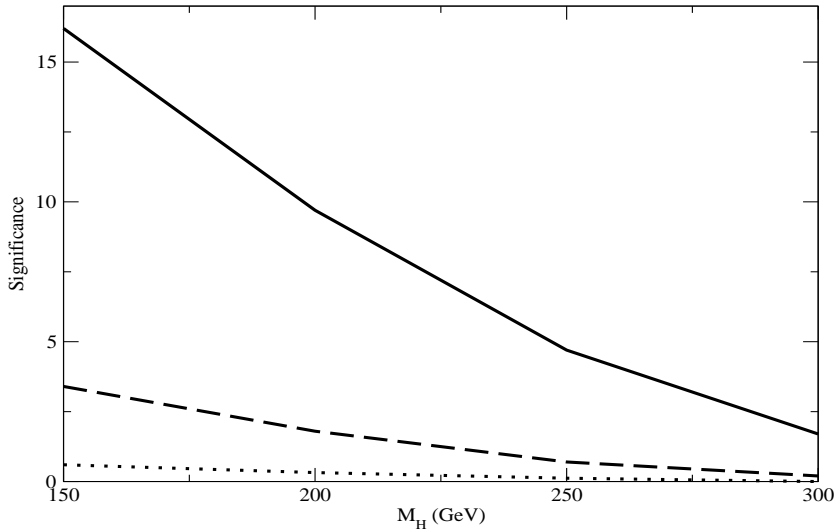,height=11cm, width=7cm, angle=-90} %
\caption{Significance of the two-Higgs signal in the Little Higgs
model with respect to the Standard model, at the LHC, assuming an
integrated luminosity of 10 $fb^{-1}$. Solid, dashed and dotted
lines correspond to $f=$ $500$, $1000$ and $2000$ GeV,
respectively. }\label{result3}
\end{figure}

We also include an analysis of the significance of the signals,
which defined as:
\begin{equation}
  \mbox{significance}=\frac{{\cal L}\cdot\sigma_{LH}-{\cal
  L}\cdot\sigma_{SM}}{\sqrt{{\cal L}\cdot\sigma_{SM}}} ,
\end{equation}
where ${\cal L}$ is the integrated luminosity of LHC and
$\sigma_{LH}, \ \sigma_{SM}$ are the two-Higgs production cross
sections for $pp$ collisions in the Little Higgs and Standard
models, respectively. Assuming a luminosity of ${\cal L}=10$
fb$^{-1}$ for the LHC, we obtain the results shown in
Fig.~\ref{result3}. Of course the significance scales as
$\sqrt{{\cal L}}$ and the figure can be easily used for larger LHC integrated luminosities. 

A word of caution is necessary at this point. It is beyond the scope of this work to include an analysis of the simulation of detector sensitivity and possible backgrounds needed for a realistic study of double Higgs production and detection in the LH model, which was done in the SM in \cite{Rainwater}. Our study is just a first step in gauging how different from the SM the LH signal for double Higgs production is.

\section{Conclusions}

The double Higgs production process probes the nature of the
electroweak symmetry breaking mechanism. This process is
intimately tied to the cancellation of quadratic divergences in
Little Higgs models. Here we have studied the reach of the LHC to
probe the LH models in this way. We found that only for relatively
small values of the energy scale $f$, of the order of $500$ to
$1000$ GeV, it is possible to distinguish meaningfully the LH from
the SM. These low values are attainable without violating the
electroweak precision limits only in models where an extra T
parity is incorporated \cite{TParity9}. On the other hand, these
results only mildly dependent on the heavy top quark mass $m_T$;
while the situation is more promising for larger values of $m_T$,
it becomes practically independent of it for $m_T$ above $2.5$
TeV.

\section*{Acknowledgments}

A.Z. and C.D. received
partial support from Fondecyt (Chile) grants No.~3020002, 1030254
7030107 and 7040059. R.R. would like to thank CNPq for partial
financial support.


\begin{thebibliography}{99}
\bibitem{discrepancy89}
We noticed that this condition was not satisfied for the values of
the couplings and masses given in Han {\it et al.},
ref.~\cite{pheno5}.
\end{thebibliography}
\end{document}